\documentclass[twoside]{article}
\usepackage{amsmath, amsfonts, amssymb}
\usepackage{graphics}
\usepackage{aat}
\usepackage{tabularx}
\usepackage{aattable}

\newcommand{\etal}{\mbox{\it{et al.}}}
\newcommand{\kms}{\mbox{km s$^{-1}$}}
\allowdisplaybreaks

\begin{document}
\thispagestyle{myfirst}

\setcounter{page}{229}

\mylabel{20}{232}

\mytitle{MASERS AND OUTFLOWS IN \\
THE W3(OH)/W3(H$_2$O) REGION}

\myauthor{A.M. SOBOLEV,$^1$ E.C. SUTTON,$^2$ D.M. CRAGG,$^3$ S.P. ELLINGSEN,$^4$
D.M. MEHRINGER,$^2$ I.I. ZINCHENKO,$^5$ A.B. OSTROVSKII,$^1$ and P.D. GODFREY$^3$}

\myadress{$^{\it 1}$Astronomical Observatory, Ural State University, Ekaterinburg 620083, Russia\\
$^{\it 2}$University of Illinois, 1002 W. Green St., Urbana, IL 61801, USA\\
$^{\it 3}$Department of Chemistry, Monash University, Clayton, Victoria 3800, Australia\\
$^{\it 4}$School of Math. and Physics, U.Tasmania, Hobart, Tasmania 7001, Australia\\
$^{\it 5}$Institute of Applied Physics, Nizhny Novgorod 603600, Russia}

\mydate{(Received December 25, 2000)}

\myabstract{Methanol masers and molecular shock tracers were observed in the W3(OH)/W3(H$_2$O) region with the BIMA array and the Onsala 20m radiotelescope. Characteristics of the outflows in the region are discussed. A model of the W3(OH) methanol maser formation region is constructed.}
\mykey{HII regions---ISM:individual (W3)---masers---radio lines:ISM}

\section{INTRODUCTION}

The OH and H$_2$O maser clusters in the W3(OH)/W3(H$_2$O) region are separated by about 5$^{\prime\prime}$ (0.05 pc at a distance of 2.2 kpc) and are associated with two young stellar objects at different stages of evolution. The OH and class II methanol masers are seen toward W3(OH) which contains an expanding ultra-compact HII region (UCHII) surrounding a young O7 star (Kawamura$\&$Masson, 1998). To its east is W3(H$_2$O), a region containing H$_2$O masers and a young stellar object known as the TW object. Both objects show outflow activity (see, e.g., descriptions of the W3(OH) champagne outflow in Keto etal.,1995 and the W3(H$_2$O) jet in Wilner etal., 1999). Situated in close proximity and embedded in the same molecular core (see, e.g., Wilson etal. 1993) these objects have rather similar velocities.

Here we report observations of molecular lines with BIMA which reveal the distibution of shock tracing molecules, in particular methanol and water, molecules which also produce maser emission. Additional observations at Onsala are used to reveal overall characteristics of the outflows. New detections of maser lines are used to construct model of the class II methanol maser formation region.

\section{OBSERVATIONS OF SHOCK TRACING MOLECULES}

\begin{figure}[t]
\myf{1}{195}{185}{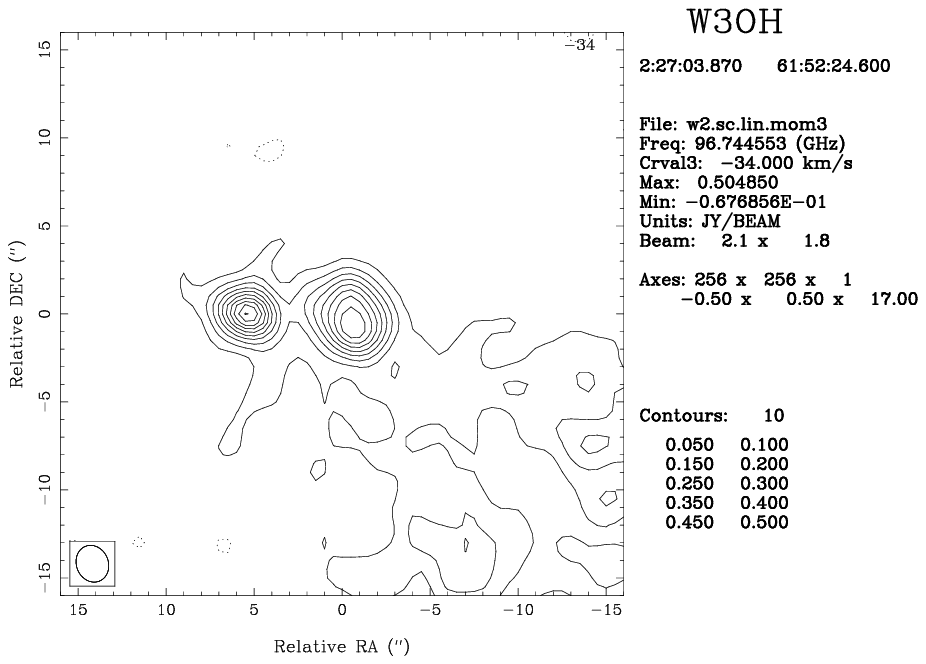}{Map of integrated flux for
blended $2_{0}-1_{0}$A$^+$ and $2_{-1}-1_{-1}$E methanol lines.
W3(OH) is in the center, W3(H$_2$O) is about 5$^{\prime\prime}$
to the east. The structure stretching to the SW from W3(H$_2$O) is
clearly seen. Contour levels correspond to multiples of 0.85 $Jy
\times \kms\ \times$ beam$^{-1}$.}
\end{figure}

Observations of the $2_k-1_k$ methanol line series, C$^{34}$S$(2-1)$, and SiO$(2-1)$ were performed with BIMA and the Onsala 20m. Additional lines, including HDO $1(1,0)-1(1,1)$, were observed with BIMA.

Both W3(OH) and W3(H$_2$O) were found emitting in the methanol lines. This indicates the presence of considerable amounts of shock processed molecular material around both objects. An important finding of our observations is the detection of an extended structure stretching to the SW from W3(H$_2$O). Emission from this structure is concentrated in the range of velocities from $-$48 to $-$44 \kms\ which is in between the characteristic velocities of W3(OH) and W3(H$_2$O).

The extended structure is well pronounced in C$^{34}$S emission as
well. The C$^{34}$S line at W3(OH) contains a clear absorption
component. So, material in an extended structure is situated in
front of W3(OH). Channel maps of C$^{34}$S emission around
W3(H$_2$O) are similar to those corresponding to a rotating disk
with the axis in an approximately $N-S$ direction.

In the BIMA results the SiO emission is located to the $SW$ of W3(H$_2$O) and is concentrated in the velocity range from $-$50 to $-$45 \kms\ . The SiO feature is less extended and is located more directly south in comparison with the extended structure seen in C$^{34}$S and methanol. Both W3(OH) and W3(H$_2$O) are not well pronounced in SiO emission. Most probably this reflects the fact that SiO emission traces the more energetic parts of the outflow than the emission of methanol and C$^{34}$S. The outflow from W3(H$_2$O) is likely to interact with the dense filament observed in NH$_3$ emission by Wilson etal. (1993).

The Onsala observations show that the W3(OH)/W3(H$_2$O) molecular
core has a size of about 1$'$ in all observed lines and shows a
$NE-SW$ asymmetry in the velocity structure.

In BIMA observations the HDO line emission is well pronounced and strongly concentrated at W3(H$_2$O). So, H$_2$O masers are situated in the part of the molecular core which is most abundant in water.

\section{OBSERVATIONS AND MODELING OF METHANOL MASERS IN W3(OH)}

Observations of nine class II methanol maser-candidate lines from the list of Sobolev, Cragg, and Godfrey (1997) were made toward W3(OH) with the BIMA interferometer. Narrow maser emission spikes at v$_{lsr}$ = $-$43.1 \kms\ are present in three of the lines: 3$_1$ -- 4$_0$ A$^+$ (107.0 GHz), 7$_2$ -- 6$_3$ A$^+$ (86.9 GHz), and 7$_2$ -- 6$_3$ A$^-$ (86.6 GHz). For all three lines the maser position is near the northern edge of the W3(OH) UCHII region. Maser emission is also seen near the southern edge at 107 GHz, but the 86 GHz lines show no evidence of this southern lobe. The spike emission in the 86 GHz 7$_2$ -- 6$_3$ A$^{\pm}$ lines is spatially compact and coincident with the bright northern component seen at 107 GHz, within an uncertainty of about 0.1 arcsec. For the remaining six lines there is no counterpart to the narrow maser spike at $-$43.1 \kms. Flux densities and upper limits for the transitions observed are given in Sutton \etal\ (2001).

The analysis which follows is done in the context of the Sobolev and Deguchi (1994) class II methanol maser model, in which methanol molecules undergo excitation to their first and second torsionally excited states by infrared radiation from warm dust. The parameters of the model are kinetic temperature ($T_{\rm kin}$), molecular hydrogen number density ($n_{\rm H}$), beaming factor ($\epsilon^{-1}$), geometrical dilution of the background free-free emission ($W_{\rm HII}$), dust temperature ($T_{\rm dust}$), and specific column density ($N_{\rm M}/\Delta V$, methanol column density in the tangential direction divided by linewidth).  We have carried out extensive new model calculations over a wide range of parameters to seek model conditions consistent with the new observations.

Fig.2 shows the model brightness temperatures for selected masers as $T_{\rm kin}$ is varied.  The ratio of brightness temperature in the 81.0 GHz $7_2$ -- $8_1$ A$^-$ transition relative to that in the 86 GHz $7_2$ -- $6_3$ A$^{\pm}$ masers is particularly sensitive to kinetic temperature, and the observed upper limit to this ratio rules out low temperature solutions. Good high temperature, high density fits are described in Sutton \etal\ (2001). In the models the calculated brightness temperature of the 107 GHz maser is about $2 \times 10^7$ K, and the implied source size is about 0.02 arcsec (40 AU at 2.2 kpc), within the range established by Slysh \etal\ (1999).  The diameter of the UCHII region is $\sim 3000$ AU, which determines the maximum maser path length, and we obtain a minimum methanol fractional abundance $10^{-6}$.  Values this large are thought to be due to grain mantle evaporation due to shocks associated with outflows or ionization fronts.

\begin{figure}[t]
\myf{2}{280}{156}{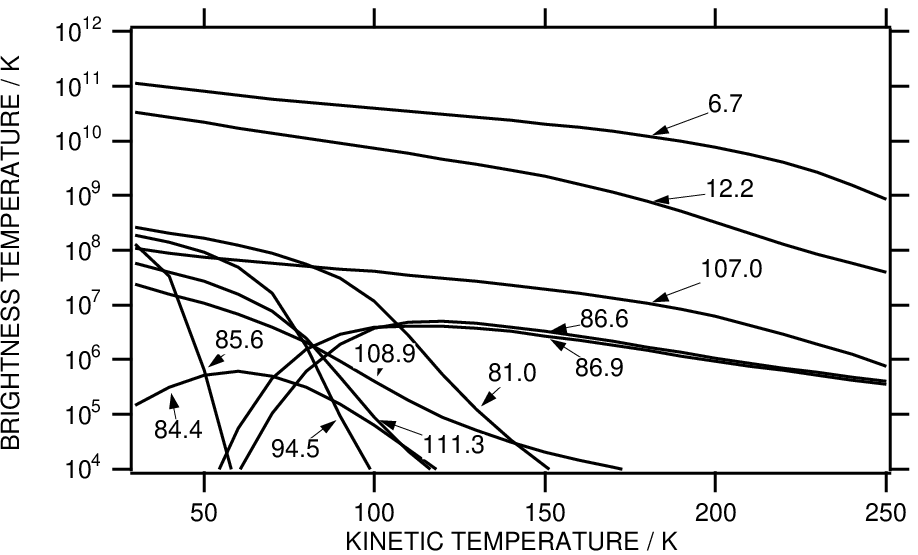}{Variation of brightness
temperature with kinetic temperature $T_{\rm kin}$ for selected
methanol masers, with $n_{\rm H} = 10^{6.8}\,\rm{cm^{-3}}$,
$N_{\rm M}/\Delta V=10^{12}\,\rm{cm^{-3}s}$, $T_{\rm dust}=175$ K,
$\epsilon^{-1}=10$, $W_{\rm HII}=0.002$.}
\end{figure}

\section{CONCLUSIONS}
The W3(OH)/W3(H$_2$O) region contains a considerable amount of shocked material mainly concentrated around W3(OH) and W3(H$_2$O), sources associated with young stellar objects. An extended structure stretching to SW from W3(H$_2$O) is detected in emission of C$^{34}$S and methanol. The extended structure and SiO emission most probably trace the outflow from the W3(H$_2$O) young stellar object.

There is a considerable amount of methanol rich gas in the immediate vicinity of the W3(OH) UCHII and in front of it.

Emission of deuterated water is concentrated near the water maser source.

From observations of nine class II methanol maser-candidate lines, we conclude that the narrow maser emission spikes arise in high density ($n_{\rm H} \approx 10^{6.8}\,\rm{cm^{-3}}$), high temperature ($T_{\rm kin} \approx 150$ K) gas, which has high methanol abundance ($>10^{-6}$) relative to molecular hydrogen.

.

Russian coauthors acknowledge support from INTAS and RBRF.

\subsection*{References}

\rf{Kawamura, J.H., and Masson, C.R. (1998)
{\it ApJ}
{\bf 509} 270.}
\rf{Keto, E.R., Welch, W.J., Reid, M.J., and Ho, P.T.P. (1995)
{\it ApJ}
{\bf 444} 765.}
\rf{Slysh, V. I., Val'tts, I. E., Kalenskii, S. V., and Larionov, G. M. (1999)
{\it Astr. Reports}
{\bf 43} 657.}
\rf{Sobolev, A. M., Cragg, D. M., and Godfrey, P. D. (1997)
{\it MNRAS}
{\bf 288} L39.}
\rf{Sobolev, A. M., and Deguchi, S. (1994)
{\it A\&A}
{\bf 291} 569.}
\rf{Sutton, E.C., Sobolev, A.M., Ellingsen, S.P., Cragg, D.M., Mehringer, D.M., Ostrovskii, A.B., and Godfrey, P.D. (2001)
{\it ApJ} in press.}
\rf{Wilner, D.J., Reid, M.J., and Menten, K.M. (1999)
{\it ApJ}
{\bf 513} 775.}
\rf{Wilson, T.L., Gaume, R.A., and Johnston, K.J. (1993)
{\it ApJ}
{\bf 402} 230.}

\end{document}